\documentclass[conference]{IEEEtran}
\IEEEoverridecommandlockouts
\usepackage{cite}
\usepackage[table]{xcolor}
\usepackage{amsmath,amssymb,amsfonts}
\usepackage{amssymb}
\usepackage{hyperref}
\usepackage{algorithmic}
\usepackage{graphicx}
\usepackage{textcomp}
\usepackage{xcolor}
\usepackage{commath}
\usepackage{multirow}
\usepackage{multicol}
\usepackage{mathtools}

\usepackage{soul,xcolor}
\usepackage{todonotes}
\setstcolor{red}

\usepackage[shortlabels]{enumitem}

\newcommand{\R}{\mathbb{R}}

\newcommand{\matx}[1]{\mathbf{#1}}
\newcommand{\vctr}[1]{\matx{#1}}
\newcommand{\tnsr}[1]{\mathcal{#1}}

\title{An Analysis of COVID-19 Knowledge Graph Construction and Applications
\thanks{Email: doflocco@davidson.edu, junyuan.lin@lmu.edu, $\{$brycewpt, ruixiaw1, trdayhzhu, rsonthal, bertozzi, branting$\}$@ucla.edu.
This work was partially supported by the AirForce Research Laboratory and DARPA under agreement number  FA8750-18-2-0066.
This work was also supported by NSF grants DMS-2027277, DMS-2027438, and the Simons Foundation Math + X Investigator Award 510776.
}
}

\author{\IEEEauthorblockN{Dominic Flocco\IEEEauthorrefmark{1},
Bryce Palmer-Toy\IEEEauthorrefmark{2},
Ruixiao Wang\IEEEauthorrefmark{2},
Hongyu Zhu\IEEEauthorrefmark{2},
Rishi Sonthalia\IEEEauthorrefmark{2},
Junyuan Lin\IEEEauthorrefmark{3}, \\
Andrea L. Bertozzi\IEEEauthorrefmark{2}
and
P. Jeffrey Brantingham\IEEEauthorrefmark{4}}
\\
	\IEEEauthorrefmark{1}Department of Mathematics and Computer Science, Davidson College, Davidson, NC 28036 \\
	\IEEEauthorrefmark{2}Department of Mathematics, University of California, Los Angeles, Los Angeles, CA 90095 \\
	\IEEEauthorrefmark{3}Department of Mathematics, Loyola Marymount University, Los Angeles, CA 90045 \\
	\IEEEauthorrefmark{4}Department of Anthropology, University of California, Los Angeles, Los Angeles, CA 90095
}

\def\BibTeX{{\rm B\kern-.05em{\sc i\kern-.025em b}\kern-.08em
    T\kern-.1667em\lower.7ex\hbox{E}\kern-.125emX}}
    
\begin{document}

\maketitle

\begin{abstract}

The construction and application of knowledge graphs have seen a rapid increase across many disciplines in recent years. Additionally, the problem of uncovering relationships between developments in the COVID-19 pandemic and social media behavior is of great interest to researchers hoping to curb the spread of the disease. In this paper we present a knowledge graph constructed from COVID-19 related tweets in the Los Angeles area, supplemented with federal and state policy announcements and disease spread statistics. By incorporating dates, topics, and events as entities, we construct a knowledge graph that describes the connections between these useful information. We use natural language processing and change point analysis to extract tweet-topic, tweet-date, and event-date relations. Further analysis on the constructed knowledge graph provides insight into how tweets reflect public sentiments towards COVID-19 related topics and how changes in these sentiments correlate with real-world events.

\end{abstract}
\section{Introduction}

\begin{figure*}
    \centering
    \includegraphics[width = 0.8\linewidth]{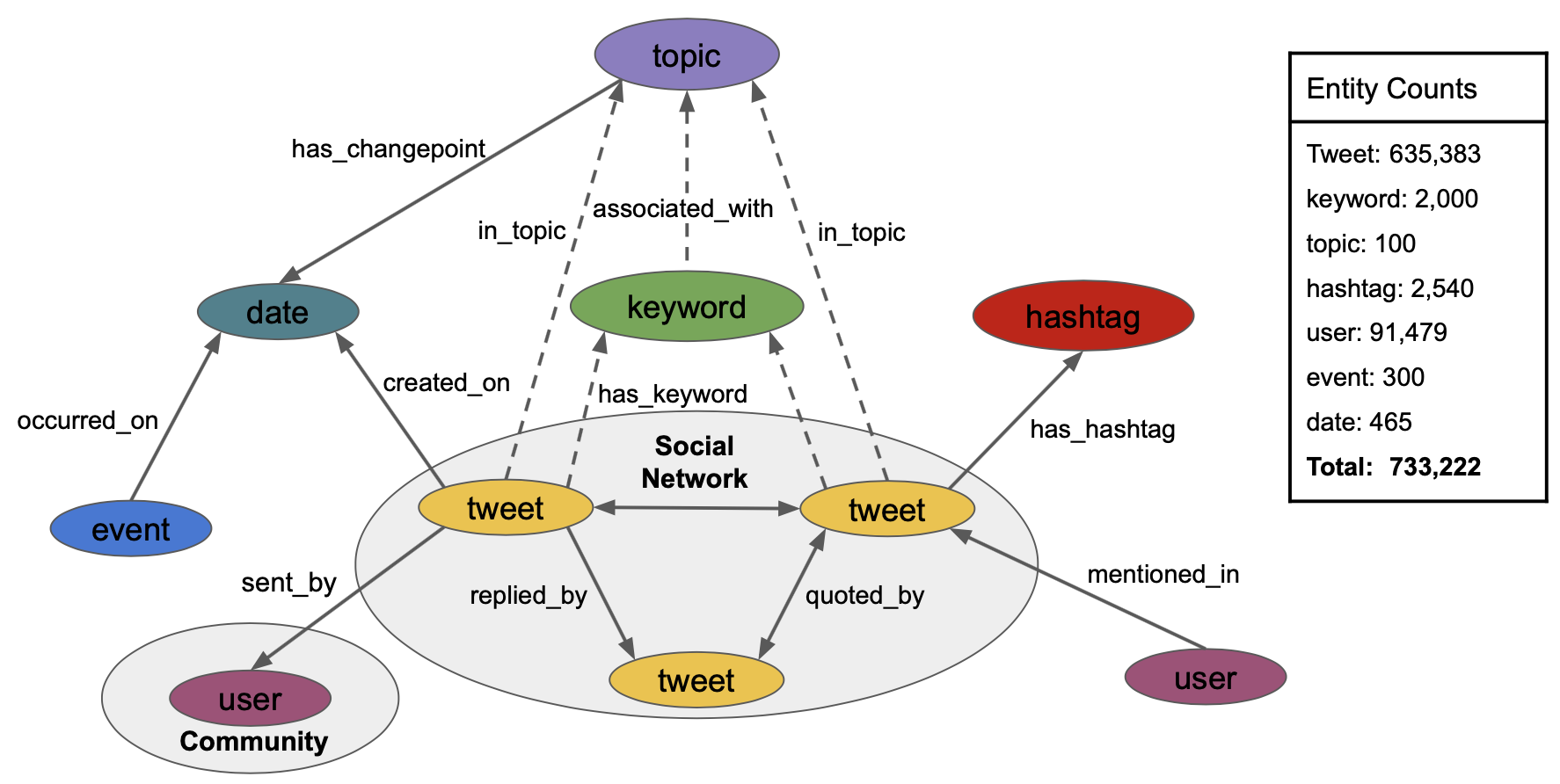}
    \caption{General structure of knowledge graph with an underlying social network of tweets at the center and external information and tweet attributes surrounding the network, where dashed edges indicate an edge weighting $w$. The tweet, hashtag and user entities are objects on extracted from Twitter; and the topic and keyword entities are generated by the NMF topic model. Event entities are complied from various news sources and reflect local and federal COVID-19 policy announcements, guideline updates, case count milestones and vaccination news. Each day spanned by the dataset is represented as a date entity.}
    \label{fig:KG structure}
\end{figure*}

The past decade has seen a rapid increase in Knowledge Graph construction and analysis due to its vast industrial and scientific applications \cite{Zou_2020}. Knowledge graphs use a directed graph structure to store and represent semantically structured information, granting it many unique strengths over other methods of data representation. Researchers can analyze relations between data, manipulate and expand the database with ease, and quickly perform navigational operations. Nodes in the graph represent objects and often store auxiliary attributes about them, and edges represent relations between objects. Moreover, techniques exist to embed the structure in vector spaces, allowing semantic analysis, knowledge classification and conventional graph learning methods. For a detailed description of common knowledge graph practices, applications and techniques, see \cite{Zou_2020, Hogan2020, Ji2020}. A well-known knowledge graph is the Google Knowledge Graph \cite{Singhal}. Other notable examples are DBpedia\cite{Lehmann}, Freebase \cite{Bollacker2007}, Wikidata \cite{Vrande2014} and YAGO \cite{Suchanek}, large open source databases of public data. 


To answer questions related to the spread of COVID-19 and the evolution of the sentiment of people in relation to COVID-19, we construct a knowledge graph using Twitter data from Los Angeles, as well as federal and state policy announcements, and disease spread statistics.\footnote{https://github.com/dominicflocco/Twitter-COVID-19-Knowledge-Graph} Since the onset of the novel coronavirus pandemic, researchers from a breadth of academic disciplines have been interested in tracking the interaction between developments in the pandemic and social media use. Specifically, they would like to understand the interaction between world events, COVID-19 policy announcements, and the sentiment of people. 
A knowledge graph is well-suited for this task, as it can accurately represent the content of social media data and news announcements while preserving the underlying structure of the social network and the relationships between entities. 

Our analysis of this dataset shows that public opinions on Twitter closely reflects real-world events suggesting that content on social media is strongly connected to real-world events. By topic modelling, we extract hotly discussed topics on Twitter with associated keywords. Through sentiment analysis and change point detection, we relate temporal changes in the sentiment in each topic with important dates of real-world events. By incorporating COVID-19 policies, world events, tweets, topics, and other entities into the graph, and using edges to represent their immediate relations, one can detect relations between any two entities through a walk in the graph. We can use node similarity to examine how closely two entities are related, and link prediction to predict or negate the existence of a relation between two entities. We also classify the topics into groups through community detection. Thus, we are able to relate topics that are frequently mentioned at the same time or by same group of users.

The recent work TweetsCov19 \cite{Dimitrov2020} also compiles Twitter data into a knowledge graph to understand the spread of COVID-19 with some limitations. Specifically, it does not include topic modelling, policy data, and diseases statistics. Additionally, our knowledge graphs contains more recent Twitter data, reflecting newer developments in the pandemic such as vaccines and variants. There are other knowledge graphs that do not use social media data to  analyze trends in academic research related to the coronavirus pandemic, such as \cite{Domingo2020}, \cite{Michel2020} and \cite{Wang2020}. 
Additionally, recent work uses social media data to model Twitter discourse \cite{Wicke2021, Tsao2021, sha2020dynamic}, to understand public sentiment \cite{Feras2020, Lyu2021}, and to perform thematic analysis\cite{Thelwall2020}. However, such work does not use a knowledge graph structure.

\subsection{Notation}
We define a knowledge graph as a weighted and directed edge-labelled graph $G = (E,R,\Delta)$, where $E$ is a set of entities that represent the set of nodes, $R$ is a set of relation types that dictate edge labels, and $\Delta\subset E\times E\times R\times\R$. Every triple in $\Delta$ is a positive fact, denoted as $\xi=(h,t,r,w)\in\Delta$, where $h\in E$ and $t\in E$ are the head and tail nodes of the fact, $r\in R$ is the edge type that connects them, and $w$ is the weight of the edge. When the weight is unnecessary, we sometimes only write $(h,t,r)$ for an edge. The graph therefore has $n = |E|$ vertices, $|\Delta|$ edges, and $|R|$ relation types.
\section{Pre-processing}
Our dataset consists of more than 7 million tweets from March 11, 2020 through June 17, 2021 authored by accounts with a geography tag in Los Angeles County and tweets geotagged within this region.

\begin{figure*}[h!]
\centering
\includegraphics[width = 0.8\linewidth]{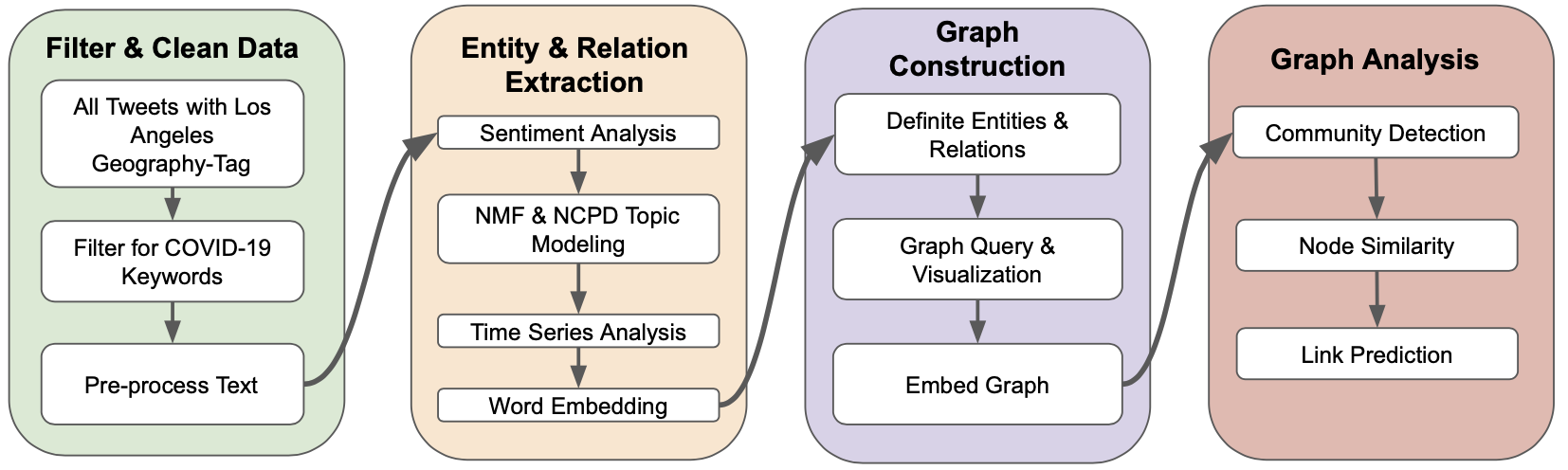}
\caption{Overview of the research process used to create and analyze the knowledge graph of interest. The filtering and cleaning of data of is outlined Section II and an overview of the models used to define relationships in the structure are presented in Section III. Graph construction and analysis techniques are then explained in detail in Sections IV and V, respectively.}
\label{fig:process}
\end{figure*}

\subsection{Filtering}
 We filter the original dataset (chosen by date and location) to include tweets pertaining to the pandemic. 
 We begin with a list of 268 COVID-19 related keywords\footnote{https://data.gesis.org/tweetscov19/keywords.txt} introduced in \cite{Dimitrov2020}. To include keywords relating to more recent COVID-19 updates like vaccines, we include keywords mentioned in \cite{Deverna2021} and the keywords used to generate the COVID-19 Twitter stream.\footnote{https://developer.twitter.com/en/docs/labs/covid19-stream/filtering-rules} Combining these three lists generated a set of 467 COVID-19 related keywords to use for filtering. 
We search the text of each tweet for these keywords. If a COVID-19 related keyword appears in the text of a tweet, that tweet and all tweets that are in the reply thread of that tweet are added to the filtered dataset. 
This allows us to include tweets that respond and react to COVID-19 information in our filtered dataset. The resulting dataset contains 635,383 tweets from 91,443 unique users.

\subsection{Cleaning the Data}
To analyze the content of our dataset and organize the information in a useful way, we pre-process the raw data before performing topic modeling, word embedding and construction of the knowledge graph. This is important given the complex language present in social media data and is crucial in extracting meaningful information about sentiment and opinion formation. We adapt the methods of \cite{Lai2016} as follows: remove non-Alphanumeric characters, user mentions, URLs, hashtags, English and Spanish stop words, and words less than three letters long. We make all characters lowercase and then tag parts of speech and lemmatize text.
These steps rid the text of unnecessary emojis, punctuation, and auxiliary information that distract from the meaning of the tweet. We store user mentions and hashtags of tweets in separate fields of the dataframe, incorporating them in the construction of the knowledge graph. This allows us to account for important information without distracting from the language of the tweet. We import stop words from the Natural Language Toolkit (NLTK) library in Python \cite{nltk} and manually expand this set with common social media slang. We perform part-of-speech tagging and lemmatization with NLTK's \texttt{WordNetLemmatizer}, yielding an abbreviated version of the text that just contains the meaningful words in the tweet. 
\begin{table}[htbp]
    \centering
    \caption{Tweet attributes stored in dataframe after pre-processing}
    
    \begin{tabular}{|l c p{5cm}|}
    \hline
         \textbf{Field}  & \textbf{Data type} &  \textbf{Description}\\ \hline
         tweet ID & \texttt{int} & Unique tweet identification number \newline generated by Twitter API \\ \hline
         created at & \texttt{str} & Date and time tweet was published \newline in UTC timezone \\ \hline
         user ID & \texttt{int} & Unqique user ID number of tweet author \newline generated by Twitter API \\ \hline
         text & \texttt{str} & Original, uncleaned text of tweet\\ \hline
         \newline cleaned text & \texttt{list str} & Cleaned and tokenized text of tweet \\ \hline
         hashtags & \texttt{list str} & List of hashtags in original 
         \newline text field denoted by \# prefix \\\hline
         mentions & \texttt{list str} & List of user mentions in original \newline text field denoted by @ prefix \\ \hline 
         in-reply-to & \texttt{int} & Tweet ID of original tweet in reply thread \\ \hline
         quoted & \texttt{int} & Tweet ID of tweet quoted by current tweet  \\ \hline
    \end{tabular}
    \label{tab:tweet-att}
    
\end{table}

In addition to storing the cleaned text, we retain other auxiliary information about each tweet extracted from our dataset stored in Twitter API format\footnote{https://developer.twitter.com/en/docs/twitter-api/v1/data-dictionary/object-model/tweet} to benefit knowledge graph construction and application. These attributes are summarized in Table \ref{tab:tweet-att}. With the processed text and attributes stored, we then perform various natural language processing tasks to provide insight into the text contents of our dataset.

\section{Natural Language Processing}
\begin{table*}[!h]
\centering 
\small
\begin{tabular}{|c|c|c|c|c|c|c|c|c|c|}\hline 
     \textbf{Topic 1} & \textbf{Topic 2} & \textbf{Topic 6} & \textbf{Topic 8} & \textbf{Topic 9} & \textbf{Topic 10} & \textbf{Topic 11} & \textbf{Topic 14} & \textbf{Topic 17} & \textbf{Topic 19}  \\\hline 
     \cellcolor{blue!20}quarantine & \cellcolor{blue!20}mask & \cellcolor{blue!20}vaccine & \cellcolor{blue!20}test & \cellcolor{blue!20}china & \cellcolor{blue!20}symptom & \cellcolor{blue!20}case & home & \cellcolor{blue!20}trump & \cellcolor{blue!20}lockdown \\ 
     chill & wear & \cellcolor{blue!20}pfizer & positive & acre & sick & \cellcolor{blue!20}report & stay & donald & another \\ 
     self & without & dose & negative & chinese & without & \cellcolor{blue!20}number & order & \cellcolor{blue!20}administration & \cellcolor{blue!20}italy \\ 
     hair & public & shot & result & magnitude &experience & total & nursing & \cellcolor{blue!20}response & whole \\ 
     bore & \cellcolor{blue!20}mandate & second & antibody & lake & mild & confirm & send & rally & full \\
     vibe & nose & \cellcolor{blue!20}moderna & site & biden & doctor & record & leave & plan & protest \\ 
     mood & damn & appointment & player & communist & fever & update & nurse & supporter & extend \\ 
     ain & protect & recieve & rapid & russia & morning & daily & safer & election & city \\ 
     wanna & store & vaccinate & testing & america & hospital & million & patient & hoax & order \\ 
     whole & walk & available & available & blame & update & increase & workout & campaign & lift \\ \hline 
    \textbf{Topic 28} & \textbf{Topic 44} & \textbf{Topic 49} & \textbf{Topic 59} & \textbf{Topic 70} & \textbf{Topic 75} & \textbf{Topic 88} & \textbf{Topic 89} & \textbf{Topic 91} & \textbf{Topic 94}  \\\hline 
    still & \cellcolor{blue!20}healthcare & hand & watch & \cellcolor{blue!20}state & \cellcolor{blue!20}social & \cellcolor{blue!20}open & happy & \cellcolor{blue!20}president & \cellcolor{blue!20}check \\ 
    despite & \cellcolor{blue!20}worker & \cellcolor{blue!20}sanitizer & \cellcolor{blue!20}movie &united & \cellcolor{blue!20}distancing & school & birthday & \cellcolor{blue!20}biden & \cellcolor{blue!20}relief \\ 
    \cellcolor{blue!20}partying & plan & wash & video & \cellcolor{blue!20}governor & distance & \cellcolor{blue!20}close & party & vice & vote \\ 
    breaker & system & smell & netflix & mandate & practice & business & \cellcolor{blue!20}celebrate & obama & bill \\ 
    flock & essential & bottle & episode & florida & medium & read & friday & force & stimulus \\ 
    spring & molina & paper & game & magnitude & guideline & store & easter & task & pass \\ 
    miami & click & shake & katie & acre & continue & high & wish & patriotic & senate \\ 
    there & frontline & wipe & youtube & lake & follow & restaurant & thanks & penny & republican \\ 
    warning & universal & toilet & porter & federal & physical & reopen & monday & fight & package \\ 
    vaccinate & nurse & soap & holy & order & maintain & everything & weekend & former & money\\ \hline 
\end{tabular}
\caption{Top keywords for 20 selected coherent topics from the 100 topics obtained from the entire dataset using NMF. Some meaningful words are highlighted.}
\label{fig:nmf_all}
\end{table*}
In this section we discuss the natural language processing (NLP) methods used to analyze sentiment, extract topics and uncover keywords from the Twitter dataset, which are then incorporated into the knowledge graph. 

\subsection{Sentiment Analysis}
Sentiment analysis on Twitter data provides insight into the user's attitude towards a particular concept, is an integral part of knowledge graph construction, and acts as the basis for time series analysis. We quantify sentiment by calculating the sentiment score $s\in [-1,1]$, where $|s|$ indicates the sentiment intensity and the sign indicates whether the sentiment is positive or negative. We use the Valence Aware Dictionary for sEntiment Reasoning (VADER) library\cite{vader} to calculate and assign the sentiment score of each tweet's original text.


To obtain numeric edge weights for tweet-to-keyword relations, we assign sentiment scores for each keyword contained in a tweet.  We implement an algorithm that yields aspect-based sentiment scores based on VADER. Given a tweet and a set of keywords appearing in the tweet, the algorithm performs the following steps. Note that for sentiment analysis, we do not use the cleaned text, but the raw text from the tweets. 

\begin{enumerate}[1.]
\item Divide a sentence into clauses, using punctuation and conjunctions as separators.\footnote{VADER detects sentiment from some punctuation and conjunctions, thus it is important where separators are placed. For our purposes, keep punctuation with their preceding clauses and the conjunctions with their following clauses.}
\item Use VADER to find the sentiment score of each clause.
\item For each aspect word, find clauses where the word appears, and assign the average score of all such clauses as the sentiment score of the relation.
\end{enumerate}

\subsection{Topic Modeling}
Topic modelling summarizes tweet contents, cluster tweets by topics, examine thematic trends, and perform additional time series analysis. The results are incorporated as topic entities and tweet-to-topic relations in our graph.
\newcommand{\ndocs}{m}
\newcommand{\nterms}{n}
\newcommand{\ntopics}{u}
\newcommand{\nslices}{s}
\newcommand{\ndates}{d}
We denote the number of documents by $\ndocs$, number of terms by $\nterms$, and number of topics by $\ntopics$. 
Given a list of terms and a corpus of tweets, we first compute the term frequency-inverse document frequency (TF-IDF) matrix, representing how often each term appears in each document \cite{tfidf}. Then we apply factorization algorithms, namely Non-negative Matrix Factorization (NMF)\cite{nmf} and Non-negative Canonical Polyadic Decomposition (NCPD)\cite{ncpd1}\cite{ncpd2}, to find a small number of hidden topics, described by their distribution over documents, terms, and time. NMF gives a summary of the topics discussed over the whole dataset, while NCPD naturally analyzes these topics over time.


Using the scikit-learn (sklearn) library \cite{sklearn}, we run NMF on the entire dataset with $\ntopics=100$ topics. Table \ref{fig:nmf_all} demonstrates a subset of our results, and suggests that NMF yields meaningful topics that reflect real-world events.

\begin{figure*}[!h]
\centerline{\includegraphics[width=0.8\textwidth]{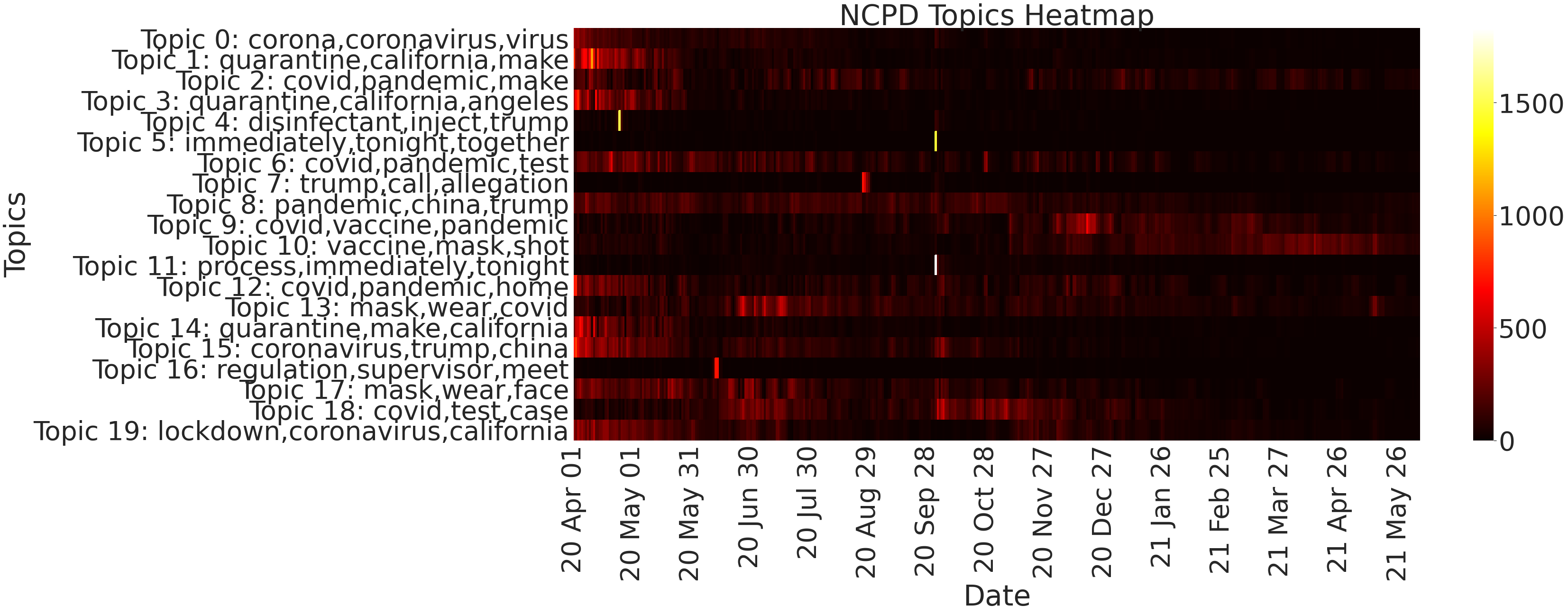}}
\caption{The evolution of 20 NCPD topics over time. To improve resolution, the data was clipped to exclude March 2020 (when an extremely high peak occurred at Topic 0). The following real-world events correspond to visible peaks in the respective topics in the heatmap: (Topic 4) Donald Trump suggested injection of disinfectants in response to COVID-19 on April 24, 2020; (Topic 16) Orange county lifted mask restrictions on June 11, 2020; (Topic 7) GOP nominated Donald Trump for 2020 President Election on August 24, 2020; (Topics 5,11) Donald and Melania Trump tested positive for COVID-19 on October 2, 2020.}
\label{fig:ncpd}
\end{figure*}

NCPD uses a data tensor $\tnsr T\in\R^{\nslices\times\ndocs\times\nterms}$, where the slice at time $i$ is the TF-IDF matrix $\tnsr T_i\in\R^{m\times n}$ of all documents at time $i$ for each $1\leq i\leq\nslices$. We divide the tweets into equal slices (continuous over time) and fix the vocabulary for the entire dataset to ensure each $\tnsr T_i$ has the same number of rows ($\ndocs$) and columns ($\nterms$), respectively. We use the TensorLy libary \cite{tensorly} to run NCPD on the whole dataset with $\nslices=306, \ndocs=2000, \ntopics=20$ (Fig~\ref{fig:ncpd}).   We observe some transient topics (e.g. topics 4, 5, 7, 11, and 16) that are discussed more frequently in a short period of time.


\subsubsection{Results}
Overall, both topic modeling approaches provide a useful summary of the dataset. The NMF topics shown in Table \ref{fig:nmf_all} are coherent and form a foundation for further analysis. The NCPD results in Fig.~\ref{fig:ncpd} not only show the popularity of recurring topics over time, but also capture localized topics corresponding to actual events. For example, Topic 16 corresponds to Orange County lifting mask restrictions. These results connect our dataset to real-world events and enable further analysis. 

\subsection{Time Series Analysis}
In order to gain insight into how COVID-19 topics and social media discourse change over time, we employ time series analysis. This also lends itself to an understanding of the juxtaposition between real-world events, public opinion and disease spread.
\subsubsection{Sentiment Change Over Time}
\begin{figure*}[h!]
\includegraphics[width = 0.7\linewidth]{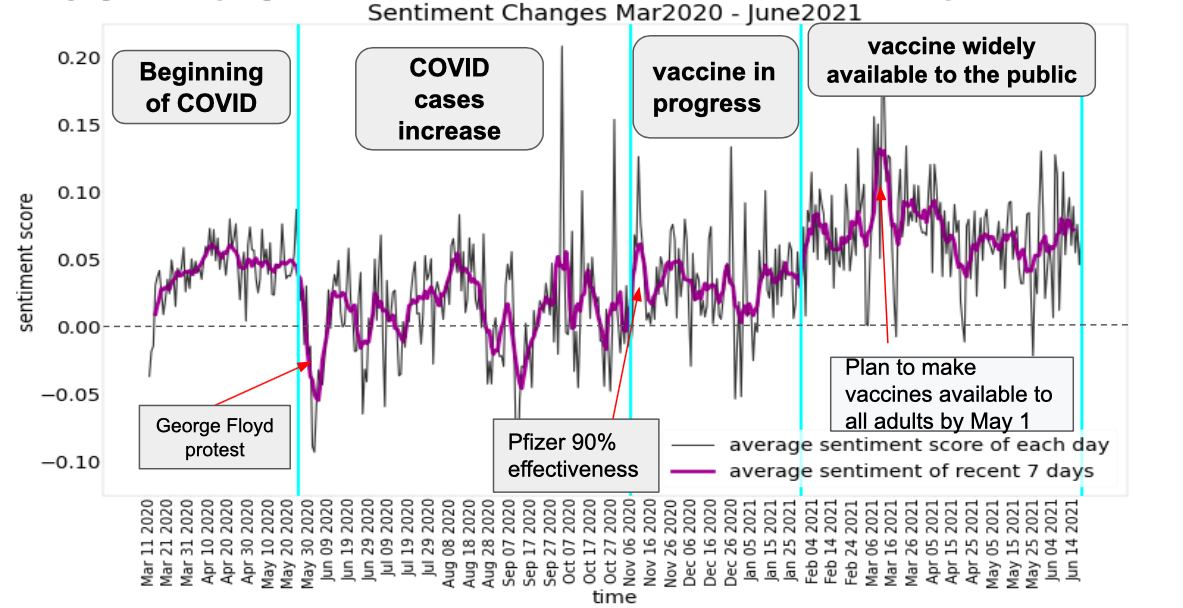}
\centering
\caption{Change point detection applied to Sentiment of all tweets, grouped by dates. Recent-7-day average sentiment is calculated to decrease the noise of oscillations. Excluding the last date which is always classified as a change point, the change points are 5/24/2020, 11/05/2020, and 1/29/2021.}
\label{fig:sentiment}
\end{figure*}

After each tweet in the dataset is assigned a sentiment score, we group tweets by date and calculate the daily average sentiment score. We observe that sentiment around COVID-19 increases when the the COVID-19 vaccine became widely available to the public around April 2021, as shown in Fig~\ref{fig:sentiment}. 

\subsubsection{Peak and Change Point Detection}

\paragraph{Volume of Tweets}
We perform a soft clustering using the 100 NMF topics such that each tweet belongs to at least one topic. Then, we plot tweet volume over time for all topics and align important events with dates at which peaks occur through analyzing text and news events. 
\begin{figure*}[h!]
\includegraphics[width = 0.9\linewidth]{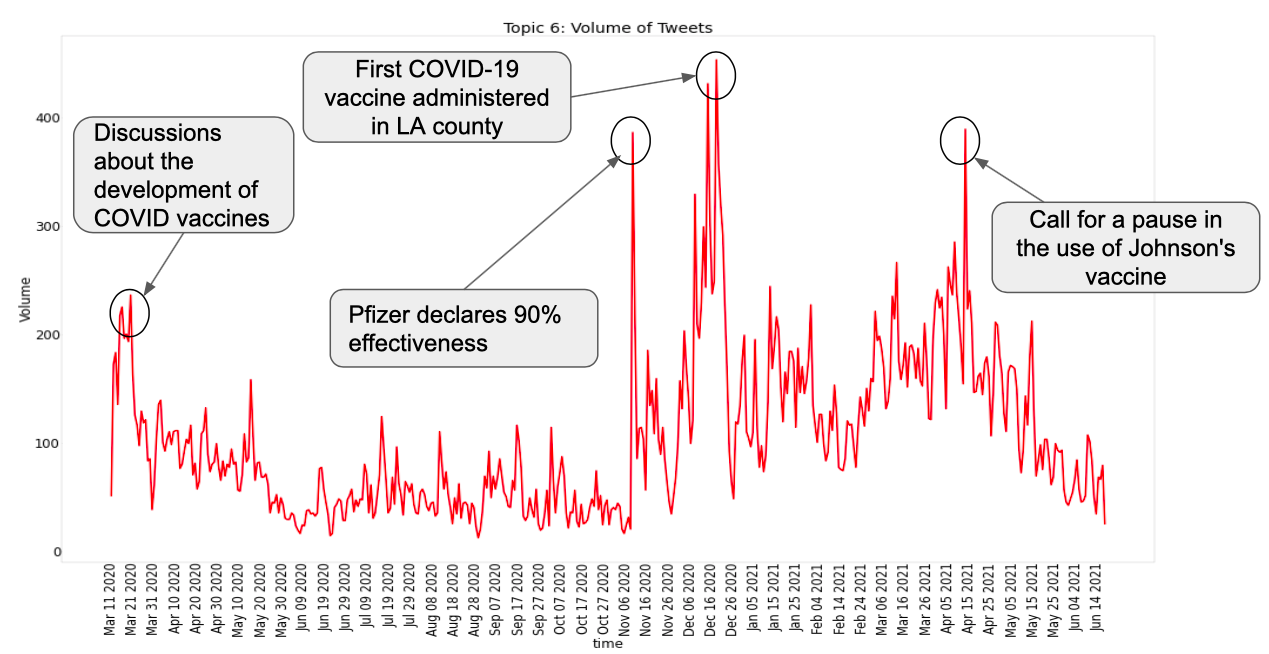}
\centering
\caption{Tweet volume of topic 6 per day. There are 4 sharp peaks on 03/21/2020, 11/09/2020, 12/18/2020, 04/13/2021, respectively, from left to right.}
\label{fig:t6vol}
\end{figure*}

Figure \ref{fig:t6vol} shows the volume of tweets in topic 6, (a vaccine topic see Table~\ref{fig:nmf_all}).
We find that peaks on 03/21/2020, 11/09/2020, 12/18/2020, 04/13/2021 correspond to considerable discussions about the development of COVID-19 vaccines: Pfizer declaring 90\% effectiveness of its vaccine, first COVID-19 vaccine administered in LA county, and the production halt of Johnson \& Johnson's vaccines, respectively. These four events are highly related to the keywords of topic 6 extracted from the NMF model. By detecting peaks in the tweet volume of each topic, we are able to identify dates when important events related to that topic take place.

\paragraph{Change Point Detection on Sentiment Change}

\mbox{}\par 

Change point detection (CPD) \cite{killick} helps detect abrupt changes in time series data and transitions between states, finding a number of change points $\tau = (\tau_{1}, ..., \tau_{m},\tau_{m+1}=n)$ that minimize a cost function (which measures changes) in an ordered sequence of data $y_{1:n} = (y_{1}, ..., y_{n})$. We use the Pruned Exact Linear Time (PELT)\cite{killick} to perform CPD as it maintains low computational costs while preserving accuracy. We adjusted the hyper-parameter of the penalty function as follows: First, we select a random topic and find important dates associated to that topic (using real world news data). Then we update the hyper-parameter so that the change points detected by PELT are approximate to the actual important dates and use the resulting value in the model.

\begin{figure*}[h!]
\includegraphics[width=0.9\linewidth]{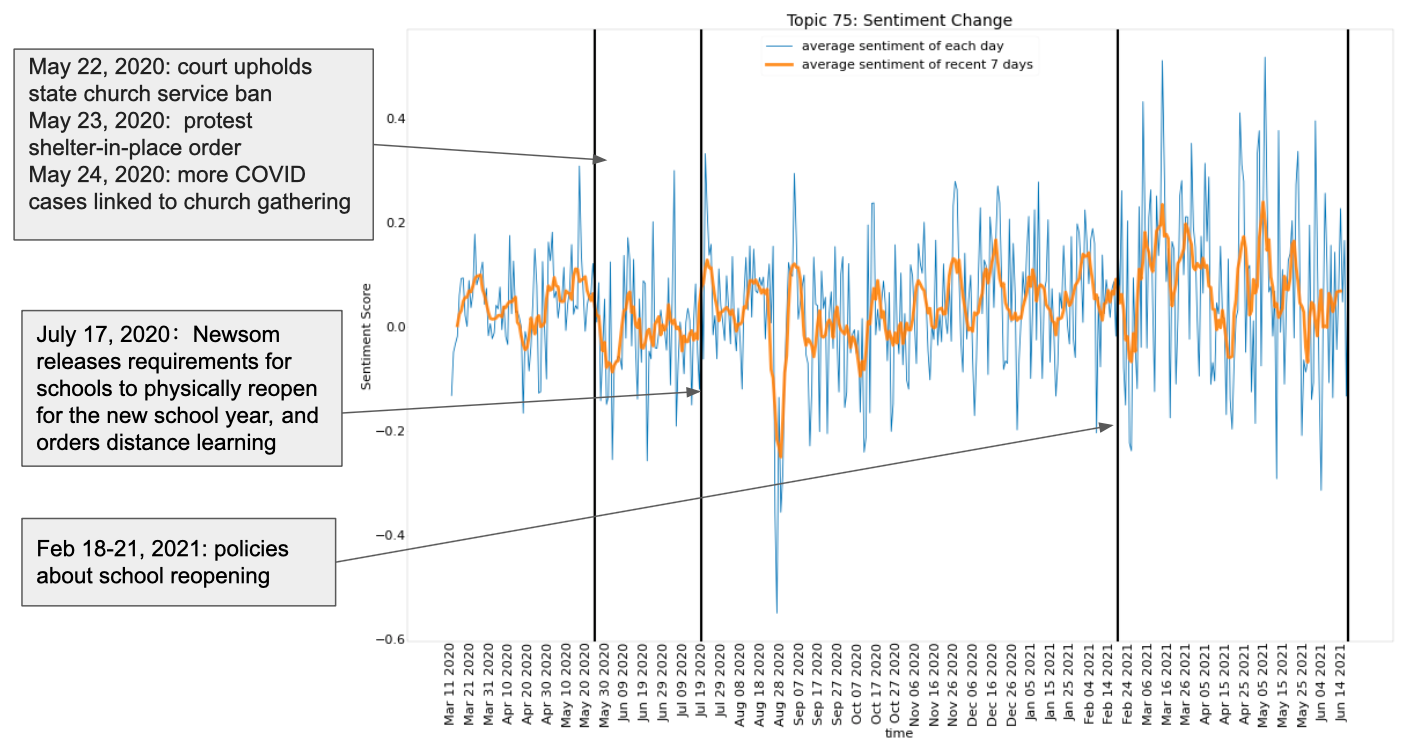}
\centering
\caption{Change point detection applied on sentiment of tweets in topic 75. Recent-7-day average sentiment is calculated to decrease the noise of oscillation. }
\label{fig:t75}
\end{figure*}

We apply change point detection to the daily average sentiment data from March 11, 2020 to June 17, 2021 to investigate how the sentiment change of tweets reflect real-world events. We present CPD results of Topic 75 in Figure \ref{fig:t75}. CPD was successful in detecting important dates associated with social distancing guidelines. However, not all 100 topics have meaningful change points that relate to real-world events. We believe these results could be improved by increasing sentiment score accuracy. It is generally difficult to accurately classify the sentiment of social media language due to abbreviations, sarcasm, irony, etc. Another possibility is that our Twitter dataset may not reflect the accurate sentiment change of a specific topic because of the inadequate tweet volumes of some NMF topics and the limitation of our dataset to only the Los Angeles area. 

These change point will be important relations in our knowledge graph. Specifically, our graph has topics and dates as entities in the graph. Hence the change points will be relations connecting a topic to the date. We will shall eventually use these our graph to predict new relations of this type. 
 
\subsection{Word Embedding}

Word embedding models allow us to measure semantic similarity between words quantitatively as the cosine similarity of the embedded vectors. With word embedding, for each tweet we can identify a word from its text and a keyword from its associated NMF topic such that the similarity between the word and the keyword is maximized. Then we add an edge between the keyword and tweet, and use the similarity value as the edge's weight.  

We use the word embedding model glove-twitter-200\cite{glove}, which is pre-trained on a corpus of common words from Twitter in 2017. We also implement Word2Vec with the Python package gensim\cite{gensim} to train another word embedding model with the words from the dataset to incorporate a more up to date and relevant dictionary. Then we combine the two models, so new words that came up after glove-twitter-200 was trained, such as "COVID-19", are also included in our model. 
\section{Graph Construction}

\indent We combine the results from our topic modeling and word embedding models with the Twitter dataset and other external information to build the knowledge graph. It represents both complex interactions in Twitter space and the important attributes of the COVID-19 pandemic around which conversation revolves. The graph is designed to simplify analyzing and understanding the interactions between the public, federal and local policymakers and the coronavirus pandemic. 

\subsection{Relations and Entities}

\indent The knowledge graph $G$ consists of seven entity types denoted by the set $E$ and 11 relation types denoted by the set $R$, which are summarized in Figure \ref{fig:KG structure}. Seven of these relations are defined by interactions and attributes in Twitter, and the remaining four are defined by the NLP models defined in Section III. We also present the general structure of our knowledge graph in Figure \ref{fig:KG structure}, where relations with an asterisk denote weighted edge attributes. The edge weight $w$ of a triplet $(h,t,r)$, where $h, t\in E$ and $r\in R$, intuitively represents the plausibility of a fact and may be thought of as the \textit{strength} of the connection between two entities.\\ 
\indent At the center of the knowledge graph is a social network of tweets derived from our original dataset. By connecting tweets by their interactions on Twitter we can represent the flow of information: this is useful in understanding how COVID-19 discourse is shaped by external influences and the ways in which concepts are related within the Twitter domain. We incorporate both information extracted from NLP techniques and temporal and event data to better understand the interaction between social media, real-world events, and the spread of COVID-19. Entities like topics and keywords allow us to cluster tweets based on their semantic similarity and shared concepts. Also, hashtags and users as independent entity types encode important aspects of interaction in the social media space. This lets us further group tweets and users by similar interests and conversation topics, providing useful insight into communities, concepts, and their relationships.  \\
\indent The event and date entities incorporate external knowledge and temporal information into the graph.  Including this information gives us insight into how social media conversation and real-world events interact. In addition, we store information about the spread of the coronavirus, such as case count, new cases, death count and new deaths, as an attribute of each date entity. We can analyze many of these metrics over time, which lends itself to an examination of the evolution of sentiment, ideas and concepts over the course of the coronavirus pandemic through time series analysis, change point detection algorithms and node similarity computation. Organizing a wide-range of information from social media, public policy and disease spread into a knowledge base can be used to shed light on many over arching social scientific questions and inform policy making. 
\begin{table}[!htbp]
\centering
\caption{Graph Statistics}
\resizebox{0.485\textwidth}{!}{%
\begin{tabular}{|l|l|l|l|l|l|l|l|l|}
\hline
Relations   & Entities  & Tweets    & Users    & Hashtags & Topics & Keywords & Events & Dates \\ \hline
$5,215,237$ & $733,222$ & $635,383$ & $91,479$ & $2,540$  & $100$  & $2,000$  & $366$  & $465$ \\ \hline
\end{tabular}%
}
\label{tab:graphstat}
\end{table}

\begin{figure}
\includegraphics[width=0.5\linewidth]{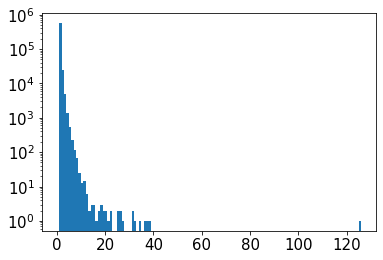}\includegraphics[width=0.5\linewidth]{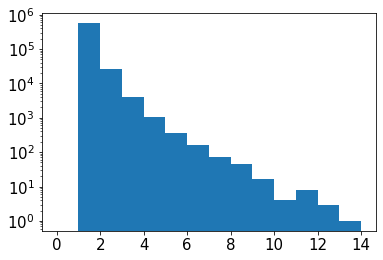}
\caption{(left) Distribution of sizes of weakly connected components in the tweet subgraph. Average: 1.076; Max: 126. (right) Distribution of longest path lengths of weakly connected
components in the tweet subgraph.  Average: 1.070; Max: 14.}
\label{fig:wcc_size}
\end{figure}

\subsubsection{Graph Statistics}
Table \ref{tab:graphstat} presents the numbers of relations, entities, and each type of entity in our knowledge graph. The knowledge graph contains sufficiently many relations so that it is weakly connected (i.e. we get a connected graph if we ignore edge direction); however, this is far from the case for the subgraph of all tweets. Fig. \ref{fig:wcc_size} shows the distribution of sizes (i.e. number of entities) and longest path lengths (the number of nodes in the longest path) in each of the $590,560$ weakly connected components of the subgraph. 
\section{Graph Applications}

\subsection{Community Detection}
We are interested in detecting communities and comparing their behavior. 
Every user or tweet is uniquely classified to one community, and the knowledge graph is divided into 31 communities, among which there are about 10 large communities. Each large community has over 70,000 tweets and 8,000 users. The 100 NMF topics are also assigned to different communities. We find that some topics are consistently grouped together into one community, similar to semantic grouping, although we did not apply any semantic analysis here. It also implies some latent relations between topics. As listed in Table \ref{tab:communities}, topics inside some communities can be summarized as a coherent, aggregated general topic.

By observing dense connections in the knowledge graph, we can detect communities in the Twitter communication network. In some communities, we find distinct popular topics being discussed during the pandemic. We also see mixed topics in some communities. For example, in the community with aggregated topic ``holidays'' and ``traveling,'' ``stimulus check'' is also discussed. We speculate that topics related to different keywords yet grouped into the same community might have hidden relations, which is worth investigation.

We used one of the most common community detection methods called the Louvain algorithm, detailed in \cite{Campigottto2014}, which is typically applied to undirected, unweighted graphs based on modularity optimization. Focusing on maximizing the density of edges inside each community, we apply the algorithm to our knowledge graph while temporarily ignoring the edge weights, edge directions, and relation types. 

\begin{table}[htbp]
    \centering
    \caption{Topics Inside Selected Communities}
    \begin{tabular}{|p{0.85cm}|p{2.cm}|p{2.1cm}|p{2.1cm}|}
    \hline
         \textbf{Comm.}  & \textbf{Number of Tweets} &  \textbf{Number of Users} &  \textbf{Aggregated Topic}\\ 
    \hline
         1  &72515 &13755 &covid policies 
 \\ \hline
         2&71805&8957&hygiene
 \\ \hline
         3&70403&8940&vaccines
\\ \hline 4&94149&11564&holidays, traveling \\
\hline

    \end{tabular}
   \label{tab:communities}    
\end{table}

\begin{figure*}[h!]
\includegraphics[width=0.8\textwidth]{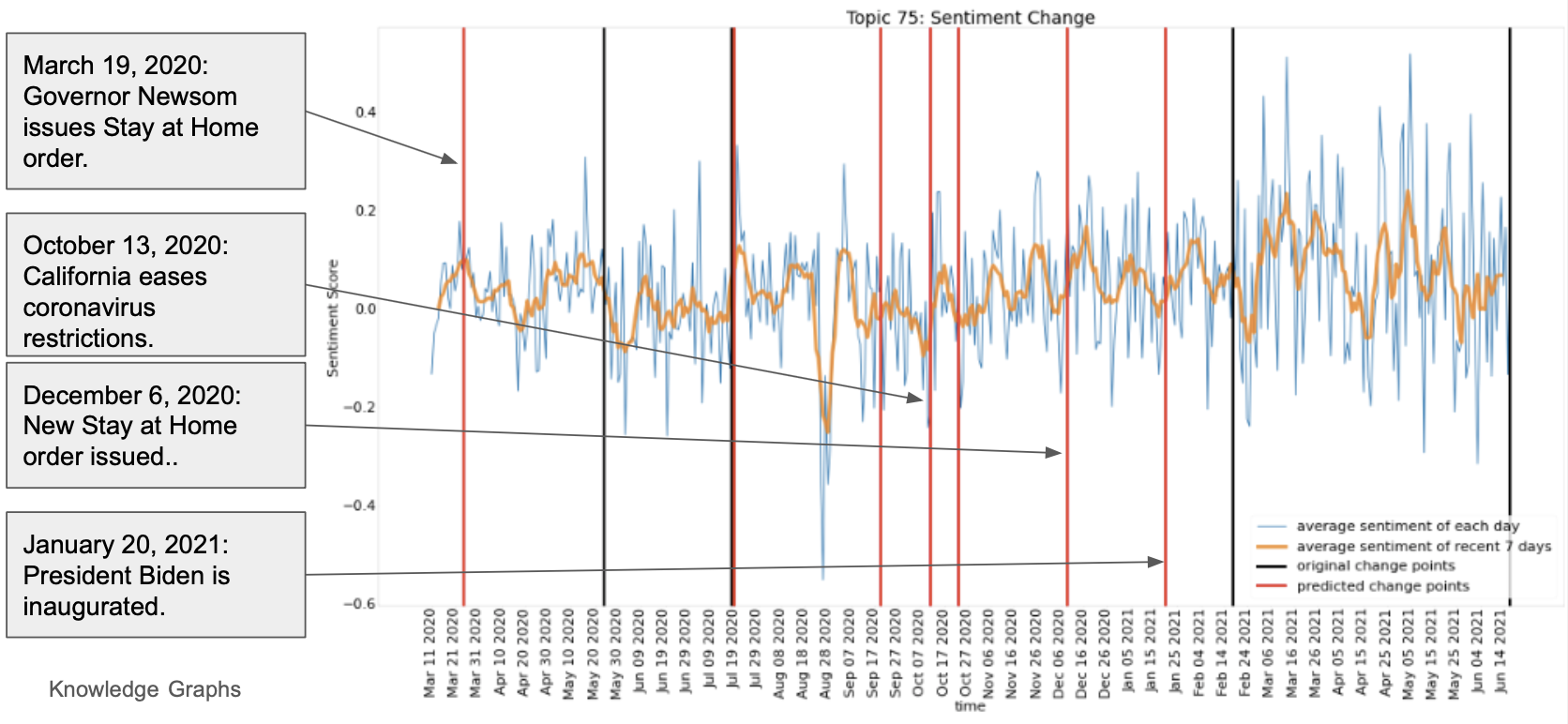}
\centering
\caption{Topic 75, about social distancing. Changepoint detection shown as black lines for the initial algorithm and red lines for the graph embedding.}
\label{fig:predchange75}
\end{figure*}

\subsection{Link Prediction}
Link prediction is the process of using a graph embedding and its scoring function to predict true relations that are not explicitly in the graph. We used an embedding to predict links for the “has\_changepoint” relationships from topics to dates in an attempt to predict sentiment changepoints. 

\subsubsection{Embedding Model}

To perform link prediction, we present a new method for embedding weighted graphs. The main idea is to embed entities and relations of a knowledge graph into continuous vector spaces while preserving the underlying structure. 
Embedding a knowledge graph is a supervised learning task that represents entities and relations as vectors and defines a scoring function on each fact to measure its plausibility. We learn an embedding by minimizing the score of observed facts and maximizing the score of \textit{negative facts}, triplets that do not appear in the knowledge base. A typical knowledge graph embedding technique consists of three steps: (i) initializing entity and relation vectors, (ii) defining a scoring function, and (iii) learning an optimal relation and entity embedding. 

We adapt a translational distance model TransD \cite{Ji2015}, measuring the plausibility of a fact as the distance between two entities after some translation is carried out by a relation. TransD uses dynamic mapping matrices to translate embeddings from the entity space to the relation space. TransD is well-suited for embedding multi-relational graphs with many types of relations and entities, so it is a good fit for our graph. For example, by defining relation-specific vector spaces for each relation, our TransD embedding reflects the semantic difference between \textit{has\_keyword} and \textit{occured\_on} relations.

\begin{figure}
\centering
\includegraphics[width=80mm]{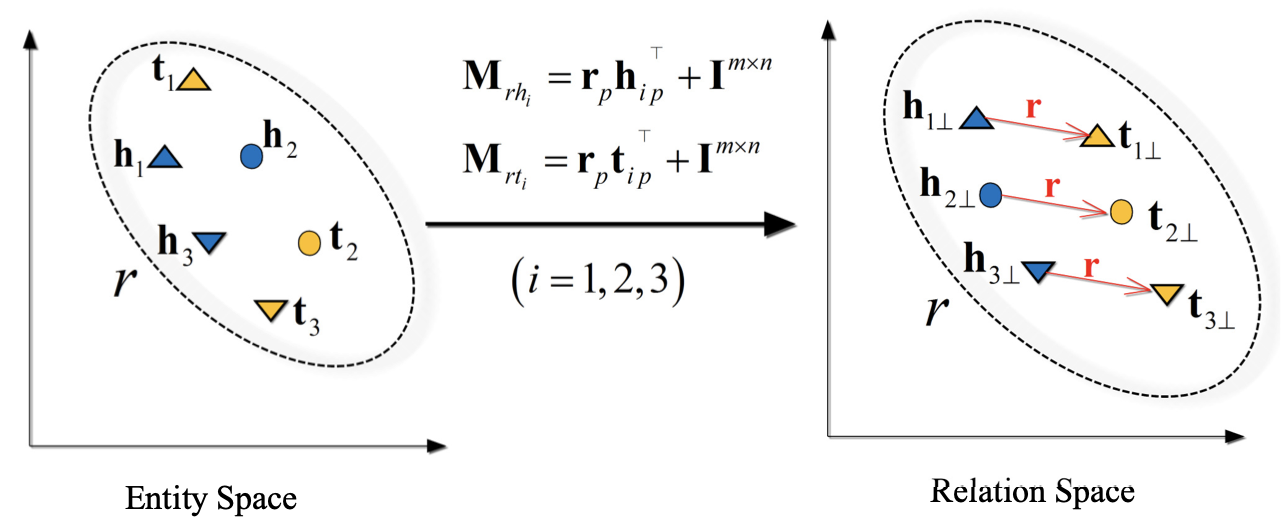}

\caption{Illustration of translation step in TransD mapping facts $(\textbf{h}, \textbf{t}, \textbf{r})$ from the entity space to the relation space via matrices $\textbf{M}_{rh}$ and $\textbf{M}_{rt}$ \cite{Ji2015}}.
\label{fig:TransD} 
\end{figure}

We begin by defining entity and relation vector spaces $\mathbb{E}\in \R^n$ and $\mathbb{P}\in \R^m$, with dimension $n$ and $m$, respectively. Given a golden triplet $\xi = (h,t,r)\in \Delta$, we define embedding vectors $\mathbf{h}, \mathbf{h}_p, \mathbf{t},\mathbf{t}_p \in \R^n$ and $\mathbf{r}, \mathbf{r}_p \R^m$, where $\cdot_p$ represents a projection vector. For each triplet $\xi $ we define two mapping matrices $\textbf{M}_{rh}, \textbf{M}_{rt} \in \R^{mxn}$ given by 
\begin{equation}
    \textbf{M}_{rh} = \textbf{r}_p\textbf{h}_p^\top + \textbf{I}^{m\times n},
    \quad
    \textbf{M}_{rt} = \textbf{r}_p\textbf{t}_p^\top + \textbf{I}^{m\times n},
    \label{eq:Mrt}
\end{equation}
which project entities from the entity space to the relation space. We then use these mapping matrices to compute the projected vectors as 
\begin{equation}
    \textbf{h}_\bot = \textbf{M}_{rh}\textbf{h}, \quad \textbf{t}_\bot = \textbf{M}_{rt}\textbf{t}.
    \label{eq:e-perp}
\end{equation}
The entity projection vectors are defined in terms of the mapping matrices, specific to the relation type. This ensures that a fact with relation $r$ is embedded into a vector space unique to that relation. An illustration of the translation via (\ref{eq:e-perp}) is shown in Fig.~\ref{fig:TransD}\cite{Ji2015}. We then use the following scoring function that incorporates these projection vectors: 
\begin{equation}
    f_r(\textbf{h}, \textbf{t}) = - \norm{\textbf{h}_\bot + \textbf{r} - \textbf{t}_\bot}_2^2
    \label{eq:TransD-score}
\end{equation}

\begin{figure*}[h!]
\includegraphics[width=0.8\textwidth]{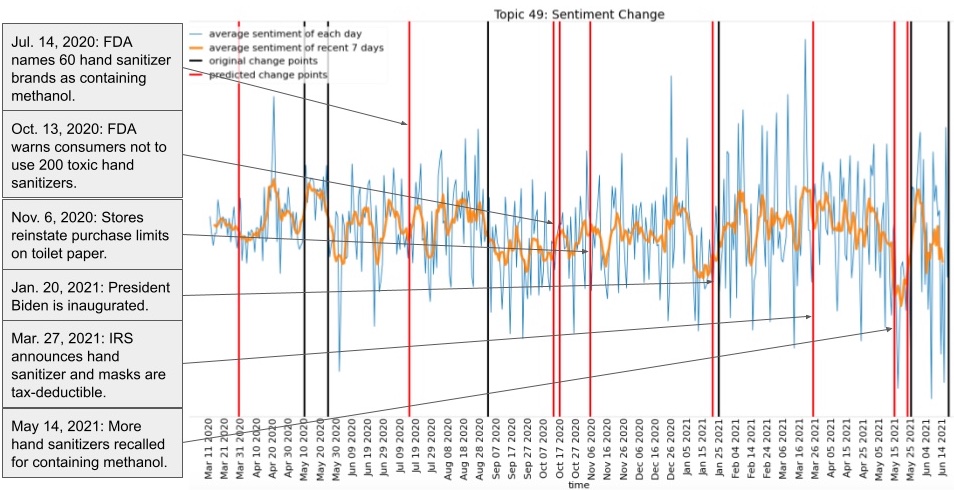}
\centering
\caption{Topic 49, about hygiene products. Change point detection shown as black lines for the initial algorithm and red lines for the graph embedding.}
\label{fig:predchange49}
\end{figure*}

\subsubsection{Incorporating Numeric Edge Weights}

Conventional translational embedding models embed only the facts in the knowledge base and ignore auxiliary information like node attributes and edge weights. We modify the conventional TransD scoring layer to incorporate edge weights into the embedding by combining it with a recently developed technique called FocusE, first introduced in \cite{Pai2021}, which improves the predictive abilities of our embedding. To our knowledge, we are the first group to combine TransD with FocusE. \\ 
\indent FocusE is an add-on layer in the learning cycle that is inserted between the scoring and loss layers and can be used with many conventional knowledge graph embedding models. The fundamental assumption it makes is that a numeric edge weight between two nodes represents the \textit{plausibility} that a link between them exists. This fits our graph because tweet-to-keyword, tweet-to-topic and keyword-to-topic relations have weights that are proportional to the degree to which these entities are related. \\
\indent We now extend our definition of a golden triplet to include this numeric edge weight, given by $\Delta = \{(h,t,r,w)\} \subset E \times E \times R \times \R$, where $w \in [0,1]$. Consider the TransD scoring function defined in (\ref{eq:TransD-score}). We first impose a softplus non-linearity $\sigma$ to ensure the values returned by $f_r$ are nonnegative: 
\begin{equation}
    g_r(\textbf{h}, \textbf{t}) = \sigma(f_r(\textbf{h}, \textbf{t})) = \ln(1+e^{f_r(\textbf{h}, \textbf{t})})
    \label{eq:softplus}
\end{equation}
Next, we introduce a modulating factor $\alpha \in \R$ to incorporate the numeric edge weight $w$ for a given triplet $\xi$. Defined as 
\begin{equation}
    \alpha = 
    \begin{cases*}
    \beta + (1-w)(1-\beta) \quad \text{if } \xi \in \Delta\\
    \beta + w(1-\beta) \quad \quad \quad \space \space \text{if } \xi \notin \Delta
    \end{cases*}
    \label{eq:alpha}
\end{equation}
where $\beta \in [0,1]$ measures the structural influence of the graph and is used to re-weight the triplet based on its edge weight. With $\beta = 1$ the edge weight $w$ is ignored and the model performs as a conventional embedding model. When $\beta = 0$ the edge weight is used without any modulating factor. Defining $\alpha$ in this way allows is to find a balance between the underlying structure of the graph and the influence of numeric edge weights. The FocusE layer is then defined as 
$h_r(\textbf{h}, \textbf{t}) = \alpha\cdot g_r(\textbf{h}, \textbf{t})$
\indent Incorporating these numeric edge weights is especially important in our graph because the links between topics, keywords and tweets are defined by external models, and are best estimates for the presence of a link between these entities. The introduction of the FocusE layer in our model allows us to reflect the uncertainty and variability of these models, acknowledging that they are not perfect links.
\subsubsection{Training the Embedding}
\indent We first split our set of positive triplets $\Delta$ into training and validation sets, with a 95\% to 5\% split. We adopt the methods of \cite{Wang2014} to generate the set of negative triplets $\Delta'$ by Bernoulli Negative Sampling, whose parameters are based on the number of head entities per tail entity for each relation $r$. We corrupt a positive triplet $\xi$ to obtain a corresponding negative triplet $\xi'$ by replacing either the head or tail entity. Through this sampling method, each positive triplet has a corresponding negative triplet used to compute the loss function, which we seek to minimize as the training objective: 
\begin{equation}
    L = \sum_{\xi \in \Delta} \sum_{\xi' \in \Delta'} \left[\gamma + h_r(\xi') - h_r(\xi\right)]_+
    \label{eq:L}
\end{equation}
where $h_r(\textbf{h}_\bot, \textbf{t}_\bot) = h_r(\xi)$ and $h_r(\textbf{h}_\bot', \textbf{t}_\bot') = h_r(\xi')$ and $\gamma$ is a hyperparameter that represents the margin of separation between a positive and negative triplet. Our model is trained on 15 epochs with batch size of 1024 with this training objective. 

\subsubsection{Results}

In our tests, we considered any triplet with a scoring function value in the lowest 2 percent of values for its corresponding topic to suggest a link.

Our model performed quite well at predicting changepoints in sentiment trend for topics. Figure \ref{fig:predchange75} shows original and predicted sentiment changepoints for Topic 75, on social distancing. Most of the predicted changepoints line up closely with major developments in social distancing rules or other significant related events. Figure \ref{fig:predchange49} similarly shows the original and predicted changepoints for Topic 49, on hygiene products such as hand sanitizer and toilet paper. Other topics also showed promising results. This lends credibility to our NMF topics as accurately capturing timely discussion topics.

\section{Conclusion}
In this project, we constructed and analyzed a knowledge graph of COVID-related tweets. Preprocessing and NLP summarize information into meaningful topics and allow for time series analysis. The graph structure has detectable communities, a measure of similarity between entities, allows for link prediction. Such a graph structure may allow us to answer questions of real-world importance, such as correlations between Twitter sentiment, policy changes, and spread of COVID-19; relationships between topics; clusters of tweets and/or users; and identification of misinformation.

\section*{Acknowledgments} We thank Deanna Needell and Thomas Tu for helpful conversations.

  The U.S.  Government is authorized to reproduce and distribute reprints for Governmental purposes notwithstanding any copyright notation thereon. The views and  conclusions contained herein are  those of the authors and should  not  be interpreted as necessarily representing the official policies  or endorsements, either expressed or implied, of the Air Force Research Laboratory and DARPA or the U.S. Government.


\bibliography{references}

\begin{thebibliography}{10}
\providecommand{\url}[1]{#1}
\csname url@samestyle\endcsname
\providecommand{\newblock}{\relax}
\providecommand{\bibinfo}[2]{#2}
\providecommand{\BIBentrySTDinterwordspacing}{\spaceskip=0pt\relax}
\providecommand{\BIBentryALTinterwordstretchfactor}{4}
\providecommand{\BIBentryALTinterwordspacing}{\spaceskip=\fontdimen2\font plus
\BIBentryALTinterwordstretchfactor\fontdimen3\font minus
  \fontdimen4\font\relax}
\providecommand{\BIBforeignlanguage}[2]{{%
\expandafter\ifx\csname l@#1\endcsname\relax
\typeout{** WARNING: IEEEtran.bst: No hyphenation pattern has been}%
\typeout{** loaded for the language `#1'. Using the pattern for}%
\typeout{** the default language instead.}%
\else
\language=\csname l@#1\endcsname
\fi
#2}}
\providecommand{\BIBdecl}{\relax}
\BIBdecl

\bibitem{Zou_2020}
\BIBentryALTinterwordspacing
X.~Zou, ``A survey on application of knowledge graph,'' \emph{Journal of
  Physics: Conference Series}, vol. 1487, p. 012016, mar 2020. [Online].
  Available: \url{https://doi.org/10.1088/1742-6596/1487/1/012016}
\BIBentrySTDinterwordspacing

\bibitem{Hogan2020}
A.~Hogan, E.~Blomqvist, M.~Cochez, C.~d'Amato, G.~de~Melo, C.~Guti{\'{e}}rrez,
  J.~E.~L. Gayo, S.~Kirrane, S.~Neumaier, A.~Polleres, R.~Navigli, A.~N. Ngomo,
  S.~M. Rashid, A.~Rula, L.~Schmelzeisen, J.~F. Sequeda, S.~Staab, and
  A.~Zimmermann, ``Knowledge graphs,'' 2020, https://arxiv.org/abs/2003.02320.

\bibitem{Ji2020}
S.~Ji, S.~Pan, E.~Cambria, P.~Marttinen, and P.~S. Yu, ``A survey on knowledge
  graphs: Representation, acquisition and applications,'' 2020,
  https://arxiv.org/abs/2002.00388.

\bibitem{Singhal}
\BIBentryALTinterwordspacing
A.~Singhal, ``Introducing the knowledge graph: things, not strings,''
  \emph{Google Blog}, 2012. [Online]. Available: \url{https://www.blog.google/
  products/search/introducing-knowledge-graph-things-not/}
\BIBentrySTDinterwordspacing

\bibitem{Lehmann}
J.~Lehmann, R.~Isele, M.~Jakob, A.~Jentzsch, D.~Konotokostas, P.~Mendes~N.,
  S.~Hellmann, M.~Morsey, P.~van Kleef, S.~Auer, and C.~Cizer, ``Dbpedia - a
  large-scale, multilingual knowledge base extracted from wikipedia,''
  \emph{Semantic Web Journal}, vol.~6, no.~2, pp. 167--195, 2015.

\bibitem{Bollacker2007}
K.~Bollacker, P.~Tufts, T.~Pierce, and R.~Cook, ``A platform for scalable,
  collaborative, structured information integration,'' \emph{Proc. of the Intl.
  Workshop on Information Integration on the Web}, 2007.

\bibitem{Vrande2014}
D.~Vrande\v{c}i\'{c} and M.~Kr\"{o}tzsch, ``Wikidata: A free collaborative
  knowledgebase,'' \emph{Commun. ACM}, vol.~57, no.~10, p. 78–85, Sep. 2014.

\bibitem{Suchanek}
F.~Suchanek, G.~Kasneci, and G.~Weikum, ``{YAGO}: a core of semantic
  knowledge,'' \emph{16th International World Wide Web Conference, WWW2007},
  pp. 697--706, 01 2007.

\bibitem{Dimitrov2020}
D.~Dimitrov, E.~Baran, P.~Fafalios, R.~Yu, X.~Zhu, M.~Zloch, and S.~Dietze,
  ``Tweetscov19 - {A} knowledge base of semantically annotated tweets about the
  {COVID-19} pandemic,'' 2020, https://arxiv.org/abs/2006.14492.

\bibitem{Domingo2020}
D.~Domingo-Fernández, S.~Baksi, B.~Schultz, Y.~Gadiya, R.~Karki, T.~Raschka,
  C.~Ebeling, M.~Hofmann-Apitius, and A.~T. Kodamullil, ``{COVID-19 Knowledge
  Graph: a computable, multi-modal, cause-and-effect knowledge model of
  COVID-19 pathophysiology},'' \emph{Bioinformatics}, vol.~37, no.~9, pp.
  1332--1334, 12 2020.

\bibitem{Michel2020}
F.~Michel, F.~Gandon, V.~Ah-Kane, A.~Bobasheva, E.~Cabrio, O.~Corby,
  R.~Gazzotti, A.~Giboin, S.~Marro, T.~Mayer, M.~Simon, S.~Villata, and
  M.~Winckler, ``Covid-on-the-web: Knowledge graph and services to advance
  covid-19 research,'' in \emph{The Semantic Web -- ISWC 2020}, J.~Z. Pan,
  V.~Tamma, C.~d'Amato, K.~Janowicz, B.~Fu, A.~Polleres, O.~Seneviratne, and
  L.~Kagal, Eds.\hskip 1em plus 0.5em minus 0.4em\relax Cham: Springer
  International Publishing, 2020, pp. 294--310.

\bibitem{Wang2020}
Q.~Wang, M.~Li, X.~Wang, N.~N. Parulian, G.~Han, J.~Ma, J.~Tu, Y.~Lin,
  H.~Zhang, W.~Liu, A.~Chauhan, Y.~Guan, B.~Li, R.~Li, X.~Song, H.~Ji, J.~Han,
  S.~Chang, J.~Pustejovsky, J.~Rah, D.~Liem, A.~Elsayed, M.~Palmer, C.~R. Voss,
  C.~Schneider, and B.~A. Onyshkevych, ``{COVID-19} literature knowledge graph
  construction and drug repurposing report generation,'' 2020,
  https://arxiv.org/abs/2007.00576.

\bibitem{Wicke2021}
P.~Wicke and M.~M. Bolognesi, ``Covid-19 discourse on twitter: How the topics,
  sentiments, subjectivity, and figurative frames changed over time,''
  \emph{Frontiers in Communication}, vol.~6, p.~45, 2021.

\bibitem{Tsao2021}
S.-F. Tsao, H.~Chen, T.~Tisseverasinghe, Y.~Yang, L.~Li, and Z.~A. Butt, ``What
  social media told us in the time of {COVID-19}: a scoping review,'' \emph{The
  Lancet Digital Health}, vol.~3, no.~3, pp. e175--e194, 2021.

\bibitem{sha2020dynamic}
H.~Sha, M.~A. Hasan, G.~O. Mohler, and P.~J. Brantingham, ``Dynamic topic
  modeling of the {COVID-19} {Twitter} narrative among {U.S.} governors and
  cabinet executives,'' 2020, https://arxiv.org/abs/2004.11692.

\bibitem{Feras2020}
F.~Al-Obeidat, O.~Adedugbe, A.~B. Hani, E.~Benkhelifa, and M.~Majdalawieh,
  ``{Cone-KG}: A semantic knowledge graph with news content and social context
  for studying {Covid-19} news articles on social media,'' in \emph{7th Int.
  Conf. on Soc. Net. Anal., Management and Security (SNAMS)}, 2020, pp. 1--7.

\bibitem{Lyu2021}
L.~G. Lyu~J, Han~E, ``Covid-19 vaccine–related discussion on twitter: Topic
  modeling and sentiment analysis,'' \emph{J Med Internet Res}, vol.~23, no.~6,
  2021.

\bibitem{Thelwall2020}
M.~Thelwall and S.~Thelwall, ``Retweeting for {COVID-19:} consensus building,
  information sharing, dissent, and lockdown life,'' 2020,
  https://arxiv.org/abs/2004.02793.

\bibitem{Deverna2021}
M.~DeVerna, F.~Pierri, B.~Truong, J.~Bollenbacher, D.~Axelrod, N.~Loynes,
  C.~Torres-Lugo, K.-C. Yang, F.~Menczer, and J.~Bryden, ``{CoVaxxy}: A global
  collection of {English Twitter} posts about {COVID-19} vaccines,'' 2021,
  https://arxiv.org/abs/2101.07694.

\bibitem{Lai2016}
E.~L. Lai, D.~Moyer, B.~Yuan, E.~Fox, B.~Hunter, A.~L. Bertozzi, and P.~J.
  Brantingham, ``{Topic time series analysis of microblogs},'' \emph{IMA
  Journal of Applied Mathematics}, vol.~81, no.~3, pp. 409--431, 07 2016.

\bibitem{nltk}
E.~Loper and S.~Bird, ``{NLTK}: The natural language toolkit,'' \emph{Proc. of
  the ACL Workshop on Effective Tools and Methodologies for Teaching Natural
  Language Processing and Computational Linguistics}, 2002.

\bibitem{vader}
C.~Hutto and E.~Gilbert, ``Vader: A parsimonious rule-based model for sentiment
  analysis of social media text,'' \emph{Proc. of the Int. AAAI Conf. on Web
  and Social Media}, vol.~8, no.~1, pp. 216--225, May 2014.

\bibitem{tfidf}
A.~Jalilifard, V.~F. Carid{\'{a}}, A.~Mansano, and R.~Cristo, ``Semantic
  sensitive {TF-IDF} to determine word relevance in documents,'' 2020,
  https://arxiv.org/abs/2001.09896.

\bibitem{nmf}
D.~{da Kuang}, J.~Choo, and H.~Park,
  \emph{\BIBforeignlanguage{English}{Nonnegative matrix factorization for
  interactive topic modeling and document clustering}}.\hskip 1em plus 0.5em
  minus 0.4em\relax Springer International Publishing, Jan. 2015, pp. 215--243.

\bibitem{ncpd1}
J.~Carroll and J.~Chang, ``Analysis of individual differences in
  multidimensional scaling via an n-way generalization of “eckart-young”
  decomposition,'' \emph{Psychometrika}, vol.~35, pp. 283--319, 1970.

\bibitem{ncpd2}
R.~A. Harshman, ``Foundations of the parafac procedure: Models and conditions
  for an "explanatory" multi-model factor analysis,'' 1970.

\bibitem{sklearn}
F.~Pedregosa, G.~Varoquaux, A.~Gramfort, V.~Michel, B.~Thirion, O.~Grisel,
  M.~Blondel, P.~Prettenhofer, R.~Weiss, V.~Dubourg, J.~Vanderplas, A.~Passos,
  D.~Cournapeau, M.~Brucher, M.~Perrot, and E.~Duchesnay, ``Scikit-learn:
  Machine learning in python,'' \emph{J. Mach. Learn. Res.}, vol.~12, no. null,
  p. 2825–2830, Nov. 2011.

\bibitem{tensorly}
J.~Kossaifi, Y.~Panagakis, A.~Anandkumar, and M.~Pantic, ``Tensorly: Tensor
  learning in python,'' 2018.

\bibitem{killick}
R.~Killick, P.~Fearnhead, and I.~Eckley, ``Optimal detection of changepoints
  with a linear computational cost,'' \emph{Journal of the American Statistical
  Association}, vol. 107, pp. 1590--1598, 12 2012.

\bibitem{glove}
menshikh iv, ``gensim-data,''
  \url{https://github.com/RaRe-Technologies/gensim-data}, 2018.

\bibitem{gensim}
R.~{\v R}eh{\r u}{\v r}ek and P.~Sojka,
  ``\BIBforeignlanguage{English}{{Software Framework for Topic Modelling with
  Large Corpora}},'' in \emph{\BIBforeignlanguage{English}{{Proc. of the LREC
  2010 Workshop on New Challenges for NLP Frameworks}}}.\hskip 1em plus 0.5em
  minus 0.4em\relax Valletta, Malta: ELRA, 2010, pp. 45--50.

\bibitem{Campigottto2014}
R.~Campigotto, P.~Conde-Céspedes, and J.-L. Guillaume, ``A generalized and
  adaptive method for community detection,'' 2014,
  https://arxiv.org/pdf/1406.2518.pdf.

\bibitem{Ji2015}
G.~Ji, S.~He, L.~Xu, K.~Liu, and J.~Zhao, ``Knowledge graph embedding via
  dynamic mapping matrix,'' \emph{Proc. 53rd Annual Meeting of the Assoc. for
  Comp. Linguistics and the 7th Int. Joint Conf. on Natural Language Processing
  (Volume 1: Long Papers)}, pp. 687--696, 2015.

\bibitem{Pai2021}
S.~Pai and L.~Costabello, ``Learning embeddings from knowledge graphs with
  numeric edge attributes,'' 2021, https://arxiv.org/abs/2105.08683.

\bibitem{Wang2014}
Z.~Wang, J.~Zhang, J.~Feng, and Z.~Chen, ``Knowledge graph embedding by
  translating on hyperplanes,'' \emph{Proc. AAAI}, p. 1112–1119, 2014.

\end{thebibliography}
\bibliographystyle{IEEEtran}

\end{document}